\documentclass[aps, prb, reprint, superscriptaddress]{revtex4-2}

\usepackage{graphicx} 
\usepackage{braket}
\usepackage{physics}
\usepackage{amsmath}
\usepackage{amssymb}
\usepackage{mathtools}
\usepackage{hyperref}
\usepackage[capitalize]{cleveref}
\usepackage{xcolor}

\begin{document}

\title{Odd fluids from chiral cellular automata}

\author{Andrew A. Allocca}
\affiliation{Physics Department, City College of the City University of New York, New York 10031, USA}

\author{Shiva Heidari}
\affiliation{Physics Department, City College of the City University of New York, New York 10031, USA}

\author{Thomas Iadecola}
\affiliation{Department of Physics, The Pennsylvania State University, University Park, Pennsylvania 16802, USA}
\affiliation{Center for Theory of Emergent Quantum Matter, The Pennsylvania State University, University Park, Pennsylvania 16802, USA}
\affiliation{Institute for Computational and Data Sciences, The Pennsylvania State University, University Park, Pennsylvania 16802, USA}
\affiliation{Materials Research Institute, The Pennsylvania State University, University Park, Pennsylvania 16802, USA}

\author{Armin Rahmani}
\affiliation{Department of Physics and Astronomy and Advanced Materials Science and Engineering Center, Western Washington University, Bellingham, Washington 98225, USA}

\author{Pouyan Ghaemi}
\affiliation{Physics Department, City College of the City University of New York, New York 10031, USA}
\affiliation{Physics Program, Graduate Center of City University of New York, New York 10031, USA}

\author{Sriram Ganeshan}
\affiliation{Physics Department, City College of the City University of New York, New York 10031, USA}
\affiliation{Physics Program, Graduate Center of City University of New York, New York 10031, USA}

\begin{abstract}
Cellular automata are discrete dynamical systems defined on a lattice, in which each site carries a finite set of states that evolve in time according to local deterministic rules. 
An important application of cellular automata is in lattice gas models of fluids, where the cellular automaton framework provides a particle-based microscopic description of hydrodynamic behavior. 
The macroscopic fluid equations emerge after coarse-graining over many lattice sites and time steps, offering a bottom-up route to hydrodynamics. 
A celebrated example is the Frisch-Hasslacher-Pomeau (FHP) model, an automaton defined on a two-dimensional triangular lattice that yields the two-dimensional Navier-Stokes equations upon coarse-graining. 
In this work, we construct a parity-breaking generalization of the FHP model through two modifications: introducing chiral two-body collision rules and systematically rotating particle velocities to mimic the effect of a background magnetic field. 
We show that this automaton yields a hydrodynamic model with odd viscosity, a transverse transport coefficient that is a hallmark of odd fluids. 
We verify the analytical transport coefficients using Poiseuille-flow simulations of the chiral FHP automaton. 
Our results demonstrate that the chiral automaton introduced here provides a bridge between microscopic parity-breaking scattering processes and macroscopic odd-fluid hydrodynamics.
\end{abstract}

\maketitle

\section{Introduction}

Broken spatial parity is a widespread feature of physical systems whose microscopic constituents acquire an intrinsic handedness through their mutual interactions, geometry, dynamics, or coupling to external fields. 
Such chirality appears in both engineered materials and naturally occurring systems, ranging from synthetic active particles and metamaterials~\cite{Banerjee2017, Soni2019, Scheibner2020, Fruchart2023, Nassar2020, Hargus2021, Srivastava2024} to biological systems~\cite{Tan2022}. 
Parity-breaking effects have also been directly observed in graphene through magnetic-field-induced Hall viscosity~\cite{Berdyugin2019}, while special Fermi-surface geometries have been theoretically shown to generate chiral responses in delafossites~\cite{Cook2019}.
A central question is how local chiral properties encoded in microscopic constituents or interactions manifest themselves at larger scales as emergent collective phenomena.
Understanding this connection between microscopic handedness and macroscopic or mesoscopic dynamics has become an active research direction. 
Microscopic models in which chirality arises from collisions and geometry have been systematically coarse-grained using Boltzmann kinetic theory, leading to explicit expressions for the associated odd transport coefficients~\cite{Han2021, Fruchart2022, Eren2025}, and theories for odd transport in electron hydrodynamic systems have also been developed~\cite{Scaffidi2017,Pellegrino2017, Cook2019}.

In conventional fluids described by the Navier-Stokes equations, this micro-to-macro bridge was established using rule-based lattice gas automata~\cite{Kadanoff1968,Hardy1972,Hardy1973,Hardy1976}. 
Lattice gas automata consist of classical particles that move and scatter on a lattice in discrete time according to a handful of fundamental rules, thereby allowing fine control over microscopic symmetries and conservation laws. 
The relative simplicity of lattice automata allows large systems to be easily simulated, though these simulations often suffer from strong statistical noise and limited quantitative accuracy.
This motivated the development of the lattice Boltzmann method~\cite{Chen1998} for simulations of fluid systems, where Boolean occupations are replaced by smooth distribution functions while preserving the same streaming–collision structure. 
For our purposes, however, the Boolean lattice formulation is especially valuable, as it is very well suited for encoding or breaking certain symmetries in the microscopic dynamics of the system.
Furthermore, coarse-graining to obtain the corresponding hydrodynamic theory is simple, so the connection between dynamics across scales is readily apparent. 

One of the most widely studied automata is the Frisch-Hasslacher-Pomeau (FHP) model, describing particles in a 2d triangular lattice interacting via 2- and 3-body collisions~\cite{Frisch1986, FHPReview1987, Henon1987}.
The Navier-Stokes equation is known to emerge from the FHP model in the macroscopic limit, directly linking strongly-interacting scattering dynamics at the lattice scale to collective dynamics of coarse-grained average quantities~\cite{Kadanoff1987,dHumieres1987, Kadanoff1989}. 

In this work, we develop a chiral generalization of the FHP lattice gas automaton by incorporating parity-breaking two-body collision rules together with a systematic rotation of particle velocities that can be precisely related to the Lorentz force exerted by an external magnetic field. 
Although the possibility of chiral collisions was already noted in the original FHP works, the continuum hydrodynamics emerging from such parity-breaking microscopic rules has, to the best of our knowledge, not yet been analyzed. 
Our construction provides a minimal microscopic model in which chirality is introduced directly at the level of discrete particle dynamics, while retaining the automaton structure required for controlled coarse-graining into a continuum hydrodynamic theory.
More broadly, the flexibility of lattice-gas automata makes them a useful testbed for exploring a variety of microscopic chiral mechanisms and their associated macroscopic consequences.

In this paper, we explicitly demonstrate that our chiral FHP model gives rise to macroscopic fluid dynamics exhibiting odd viscosity.
Odd viscosity, also known as Hall viscosity in condensed matter physics and gyroviscosity in plasma physics~\cite{landau1987lifshitz}, is a nondissipative transverse response to shear flows and manifests as a nonzero antisymmetric component $\eta_H$ of the viscosity tensor, for which we derive exact analytical expressions in terms of the model's microscopic parameters. 
We also find that the same chiral parameters that generate $\eta_H$ suppress the ordinary shear viscosity $\eta$, thereby directly linking microscopic parity-breaking rules to the structure of viscous transport in the continuum theory.

Hall viscosity is a distinctive transport coefficient of isotropic two-dimensional systems and has been studied in a broad range of settings, from quantum Hall fluids and electron hydrodynamics in graphene and delafossites to classical chiral and active matter~\cite{Avron1995,Tokatly2006,Tokatly2007,Haldane2011,Hoyos2012,Bradlyn2012,Abanov2013,Hoyos2014,Laskin2015,Can2015,Klevtsov2015,Scaffidi2017,Pellegrino2017,Berdyugin2019,Korving1966,Scheibner2020,Banerjee2017,Soni2019,Markovich2021,Monteiro2023,Reynolds2022}. 
The broad class of systems in which odd viscosity appears has been extensively reviewed in Ref.~\cite{Fruchart2023} and references therein. 
As such, Hall viscosity is a hallmark of parity broken fluids: it encodes the leading parity-odd component of the viscous stress and provides a direct hydrodynamic signature of microscopic chirality at macroscopic scales. Thus, our work paves the way for encoding general symmetry principles directly into microscopic automaton rules, allowing the corresponding macroscopic transport coefficients to emerge naturally upon coarse-graining in a manner relevant to several experimental platforms.

The remainder of this paper is organized as follows.
In \cref{sec:automaton} we find define the automaton rules incorporating chiral scattering and an applied magnetic field, and write the Boltzmann equation including these effects. 
In \cref{sec:hydrodynamics} we perform a multiscale continuum analysis to obtain hydrodynamic equations from our microscopic automaton, which depend on fluctuations from equilibrium that we analyze in \cref{sec:fluctuations}.
In \cref{sec:viscosity} we then extract the shear and Hall viscosity components and examine their dependence on both parity-breaking effects of the model.
In \cref{sec:simulation} we numerically simulate the automaton dynamics and extract the viscosity components to compare with theory. 
In \cref{sec:discussion} we discuss the results and the outlook for future developments. 

\section{Chiral Lattice Automaton} \label{sec:automaton}
The FHP model is defined on a 2-dimensional triangular lattice. 
Each vertex is connected to its six neighbors by unit vectors $\hat{\mathbf{c}}_\ell = \left(\cos((\ell-1)\pi/3)\,\,\,\sin((\ell-1)\pi/3)\right)^T$, with $\ell = 1,\dots,6$. 
Up to six particles may reside at each lattice site, one associate with each link, with $N_\ell(\mathbf{x},t) \in\{0,1\}$ denoting the occupation of these allowed states at time $t$, where $\mathbf{x}$ denotes the position of the vertex.  
The state of the automaton is specified by the set of all occupations at a given time.
Automaton dynamics update this state in discrete time steps, with each step composed of two sub-steps: collisions and streaming. 
\begin{figure*}[!t]
    \centering
    \includegraphics[width=0.95\textwidth]{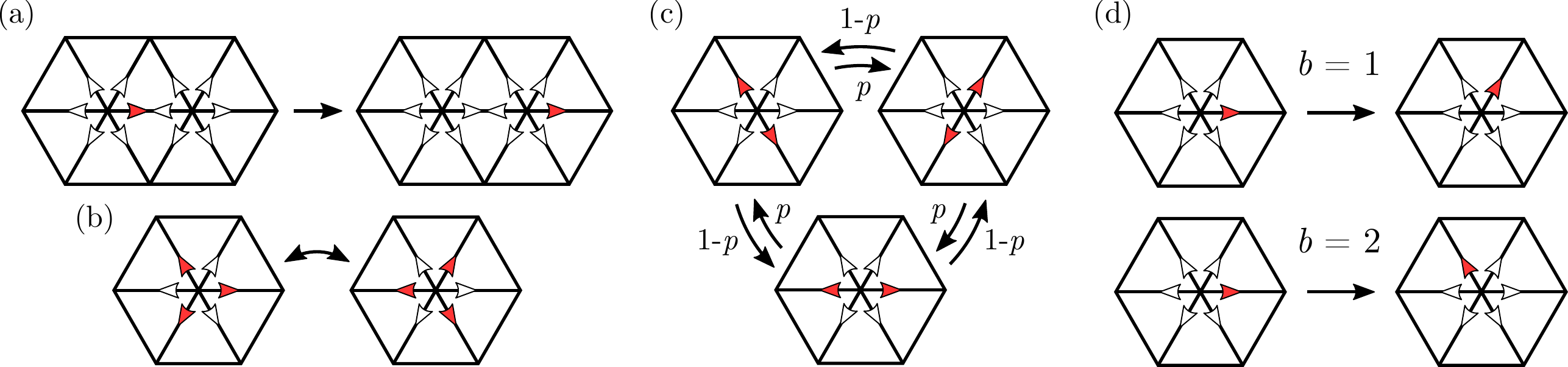}
    \caption{A graphical representation of the rules governing the time evolution of the chiral automaton.
    Particles are indicated by red arrows pointing from their associated lattice site $\mathbf{x}$ along the direction of their velocity $\ell$. 
    (a) The streaming step, moving a particle to the next lattice site in the direction of its velocity.
    (b) Three-body collisions between configurations $C_{3,i}$.
    (c) Chiral two-body collisions between configurations $C_{2,i}$; the probability for clockwise and counter-clockwise scattering are allowed to differ.
    (d) The magnetic effect demonstrated for a single particle, shifting the link index by $b$ after streaming and rotating the velocity. All configurations are rotated in this way.}
    \label{fig:automaton}
\end{figure*}

\subsection{Chiral Collision Step}
In the collision or scattering sub-step configurations of particles at each vertex scatter into other configurations according to predetermined rules.
The FHP model allows nontrivial scattering only between the 2-body and 3-body configurations with zero total momentum:
two particles on opposing links, $\ell=1,4$, $\ell=2,5$, and $\ell=3,6$, which we denote $C_{2,1}$, $C_{2,2}$, and $C_{2,3}$, or three particles on every other link, $\ell=1,3,5$ and $\ell=2,4,6$, denoted $C_{3,1}$ and $C_{3,2}$.  
Other configurations do not undergo any scattering. 

Three-body configurations $C_{3,i}$ scatter into each other with probability 1, %$\mathcal{A}_{C^{(3)}_{1},C^{(3)}_{2}} = \mathcal{A}_{C^{(3)}_{2},C^{(3)}_{1}}= 1$, 
shown graphically in \cref{fig:automaton}(b). 
This process is included to eliminate the spurious conservation of momentum along all three axes of the lattice independently and leaving just three total conserved quantities---the density and two components of momentum. 

We allow 2-body scattering in clockwise and counter-clockwise senses to have unequal probability, parametrized by a quantity $p$ as shown in \cref{fig:automaton}(c). 
The standard FHP model has $p=\tfrac{1}{2}$, but when $p\neq \tfrac{1}{2}$ the scattering is chiral and breaks both parity and time-reversal symmetry while preserving their product, in principle allowing for a nonzero Hall viscosity. 

\subsection{Streaming in a Magnetic Field} \label{subsec:2B}
The second sub-step of each time step is the streaming or propagation step.
Particles shift along the link they occupy to the next lattice site as shown in \cref{fig:automaton}(c): $N'_\ell(\mathbf{x},t)\to N_\ell(\mathbf{x}+\hat{\mathbf{c}}_\ell,t+1)$, where $N'$ indicates occupation number after scattering. 
The link index $\ell$ is thus in one-to-one correspondence with particle velocity $\hat{\mathbf{c}}_\ell$. 

The effect of a magnetic field is to rotate each particle's velocity after this free streaming, shown in \cref{fig:automaton}(d): $N_\ell(\mathbf{x},t)\to N_{\ell+b}(\mathbf{x},t)$, with $b=1,2$ corresponding to the effect of two possible nontrivial ``magnetic field strengths.'' 
For $b=1$, a rotation by $60^\circ$, an isolated particle that undergoes no scattering will return to its original state after 6 time steps, and for $b=2$, a rotation by $120^\circ$, it will return after 3 time steps---these are the lattice equivalent of cyclotron orbits. 
This effectively introduces an external magnetic force into the dynamics, breaking momentum conservation. 
However, since this rotates particle velocities together the magnitude of the total momentum $\abs{\mathbf{P}}$ is preserved, with its direction recurring every $6/b$ time steps.
We note that $b=-1, -2$ would describe a magnetic field in the opposite direction. Also, $b=3$ corresponds to flipping the velocity direction and does not capture a rotation. Other integer values of $b$ will be equivalent to $\pm 1,\pm 2$. It may appear that the construction allows for only two discrete magnetic field strengths. However, the rules can be modified to encode a continuous magnetic field, e.g., by making the rotations probabilistic.

\subsection{Lattice Boltzmann equation}
To analyze the kinetics of this automaton we exchange occupation numbers $N_\ell(\mathbf{x},t)\in\{0,1\}$ for the occupation probabilities $n_\ell(\mathbf{x},t)\in[0,1]$, which can be obtained in principle as the late-time averages of the occupation numbers or as their average over all initial states with compatible total particle number and momentum. 
Local density and momentum are
\begin{gather}
    \rho(\mathbf{x},t) = \sum_\ell n_\ell(\mathbf{x},t) \label{eq:fulldensity}\\
    \mathbf{p}(\mathbf{x},t) = \sum_\ell \hat{\mathbf{c}}_\ell \,n_\ell(\mathbf{x},t)\equiv \rho(\mathbf{x},t)\mathbf{u}(\mathbf{x},t), \label{eq:fullmomentum}
\end{gather}
with the velocity field $\mathbf{u}(\mathbf{x},t)$ defined as their ratio. 

The Boltzmann equation with the magnetic velocity rotation is
\begin{equation} \label{eq:Boltzmann}
    n_{\ell+b}(\mathbf{x}+\hat{\mathbf{c}}_\ell,t+1)-n_{\ell}(\mathbf{x},t) = \Omega_\ell\left[\{n_k(\mathbf{x},t)\}\right] 
\end{equation}
where the collision function $\Omega_\ell$ encoding the 2- and 3-body collision rules defined above is
\begin{widetext}
\begin{align} \label{eq:chiralscattering}
    \Omega_\ell\left[\{n_k\}\right] &= \Big[(1-p)\,n_{\ell-1}n_{\ell+2}(1-n_{\ell+1})(1-n_{\ell-2}) + p\,n_{\ell+1}n_{\ell-2}(1-n_{\ell-1})(1-n_{\ell+2})\Big](1-n_{\ell})(1-n_{\ell+3}) \nonumber\\
    &-n_\ell n_{\ell+3}(1-n_{\ell+1})(1-n_{\ell+2})(1-n_{\ell-1})(1-n_{\ell-2})\nonumber\\
    &+n_{\ell+1}n_{\ell-1}n_{\ell+3}(1-n_{\ell})(1-n_{\ell+2})(n_{\ell-2}) - n_\ell n_{\ell+2}n_{\ell_2}(1-n_{\ell+1})(1-n_{\ell-1})(1-n_{\ell+3}),
\end{align}
\end{widetext}
written here with $\mathbf{x}$ and $t$ suppressed.
The first and second lines give the probabilities, parametrized by $p$, for a particle to scatter onto or off of link $\ell$ by chiral 2-body collisions, and the third line gives these probabilities for 3-body collisions. 
Conservation of density and momentum by these collisions gives
\begin{equation} \label{eq:conservation}
    \sum_\ell\Omega_\ell[\left\{n_k\right\}] = \sum_{\ell}\hat{c}_\ell\,\Omega_\ell[\left\{n_k\right\}] = 0.
\end{equation} 

\subsection{Equilibrium}
The equilibrium solution of the Boltzmann equation $n^\mathrm{eq}_\ell$ is the spatially uniform and constant link occupation probability that is invariant under streaming and collisions, and is characterized by density $\rho$ and velocity $\mathbf{u}$ that are approximately constant over large regions of the system,
\begin{equation} \label{eq:uniform}
    \rho=\sum_\ell n^\mathrm{eq}_\ell \qquad
    \rho\mathbf{u} =\sum_\ell \hat{\mathbf{c}}_\ell\,n^\mathrm{eq}_\ell. 
\end{equation}
This solution provides the starting point for developing the full hydrodynamic theory below. 
The form of the equilibrium occupation is obtained by maximizing the entropy on each link subject local conservation laws (see Appendix \ref{app:equilibrium}).

In equilibrium the density and velocity are time independent, but with the magnetic rotation effect the velocity acquires a time-dependent rotation just like the total momentum: $\mathbf{u}(t+1) = R(b\pi/3)\mathbf{u}(t)$, where $R(\theta)$ is a rotation around the $z$-axis by angle $\theta$.
In the limit of low Mach number $\abs{\mathbf{u}} \ll \abs{\hat{\mathbf{c}}_\ell} = 1$ the equilibrium solution expanded up to $O(u^2)$ has the form~\cite{FHPReview1987}
\begin{equation} \label{eq:equilibrium}
    n^\mathrm{eq}_\ell = \frac{\rho}{6} + \frac{\rho}{3} \hat{c}_\ell^i u_i + \rho\,G(\rho)\,Q^{ij}_{\ell}\,u_iu_j,
\end{equation}
where $G(\rho)=\tfrac{1}{3}\tfrac{6-2\rho}{6-\rho}$ and $Q^{ij}_{\ell} \equiv \hat{c}_\ell^i\hat{c}_\ell^j - \delta^{ij}/2$. 
The coefficients of this expansion are determined by enforcing \cref{eq:uniform} while maintaining discrete rotation symmetry of the lattice. 
The factor $G(\rho)$ encodes the breaking of Galilean symmetry by the underlying lattice, which is in principle an undesirable artifact of the model~\cite{Frisch1986}. 
However, for a one-component liquid, Galilean invariance can be restored by an appropriate rescaling of time.
If $\rho$ and $\mathbf{u}$ are not completely constant through out the system but vary slowly on long length and time scales, then \cref{eq:equilibrium} captures the behavior of these quantities on scales smaller than these variations, with approximately constant density and velocity in large ``patches.''
This will be key to developing the hydrodynamic theory in \cref{sec:hydrodynamics}. 

The only effect of the magnetic rotation on \cref{eq:equilibrium} is to induce the time dependent rotation of $\mathbf{u}$.
This is clear by considering a coordinate frame that rotates by $b\pi/3$ at each time step.
In this reference frame the total momentum is constant, allowing us to define a constant co-rotating velocity $\mathbf{u}_0$. 
The derivation giving \cref{eq:equilibrium} for $b=0$ can be repeated in this rotating frame to give the same form, now in terms of $\mathbf{u}_0$. 
Transforming back to the static ``lab frame'' restores the rotation in $\mathbf{u}$, and because $n^\mathrm{eq}_\ell$ is independent of $\mathbf{x}$ this final return to the non-rotating frame has no further effect.

\section{Hydrodynamic Equations} \label{sec:hydrodynamics}

We now develop a macroscopic hydrodynamic theory of the density $\rho(\mathbf{x},t)$ and velocity $\mathbf{u}(\mathbf{x},t)$ from this chiral automaton, generalizing the approach of Ref.~\cite{FHPReview1987}. 
This procedure necessarily extends discrete $\mathbf{x}$ and $t$ to continuous quantities. 
It will also be useful to generalize the link index $\ell$ to a continuous angular degree of freedom $\theta$, with the discrete directions of the links corresponding to angles $\theta_\ell=(\ell-1)\pi/3$, so that $n_\ell(\mathbf{x},t) = \eval{n(\theta,\mathbf{x},t)}_{\theta=\theta_\ell}$ and $\hat{\mathbf{c}}_\ell=\eval{\hat{\mathbf{c}}(\theta)}_{\theta=\theta_\ell}$.
The magnetic rotation of $\ell$ to $\ell+b$ corresponds to a shift in $\theta$ by $B=b\pi/3$, which we now refer to as the magnetic field strength.
The Boltzmann equation can be rewritten as
\begin{equation} \label{eq:Boltzmann2}
    \left(\hat{T}_\ell \hat{R}_B -1\right)\eval{n(\theta,\mathbf{x},t)}_{\theta=\theta_\ell} = \Omega_\ell[\left\{n_k\right\}],
\end{equation}
where $\hat{T}_\ell = \exp\left[\partial_t+\hat{c}_\ell^i\partial_i\right]$ with $\partial_i = \partial/\partial x_i$ is the operator that advances time and translates in space by the appropriate discrete amounts and $\hat{R}_B = \exp\left[B\,\partial_\theta\right]$ is the operator that rotates the angular degree of freedom by $B$.
The angle $\theta$ is always restricted to the discrete values of the lattice system in the end. 

To obtain a hydrodynamic theory from the Boltzmann equation we employ a Chapman-Enskog expansion in multiple scales. 
The fundamental assumption of this procedure is to suppose that the system is comprised of large ``patches'' that are locally near equilibrium, with density and velocity varying between them.
Let these patches have linear dimension $1/\varepsilon\gg1$ in terms of the lattice scale, so $\varepsilon$ is a small parameter and variations in $\rho$ and $\mathbf{u}$ occur on the long length scale $\mathbf{x}_1=\varepsilon\mathbf{x}$, not at the lattice scale $\mathbf{x}$. 
Similarly, these quantities only vary appreciably on time scales much longer than the discrete time $t_0$ of the automaton dynamics, so we define the two slow scales $t_1=\varepsilon t_0$ and $t_2 = \varepsilon^2 t_0$ in order to capture both propagation of sound waves and slower diffusive/advective phenomena. 

The link occupations are expanded as $n_\ell(\mathbf{x},t) = n^0_{\ell}(\mathbf{x},t)+\varepsilon\,n^1_{\ell}(\mathbf{x},t)$, with $\ell$ generalized to $\theta$ as necessary.
The leading term $n^0$ is the local equilibrium in each large patch so it has same form as $n^\mathrm{eq}$ in \cref{eq:equilibrium}, though now parametrized by $\rho$ and $\mathbf{u}$ that are only approximately constant in the manner discussed above.
The small correction $n^1$ encodes fluctuations from this local equilibrium and provides no contribution to the local $\rho$ and $\mathbf{u}$. 
Hydrodynamic behavior ultimately arises from these fluctuations diffusing the long-wavelength inhomogeneities across patches. 
The scattering function $\Omega_\ell[\left\{n\right\}]$ must also be expanded in $\varepsilon$ due to its dependence on $n$,
\begin{equation}
    \Omega_\ell[\left\{n\right\}] \approx \Omega^0_{\ell} + \varepsilon\,\Lambda^0_{\ell\ell'}n^1_{\ell'} + \frac{\varepsilon^2}{2}\Gamma^0_{\ell\ell'\ell''}n^1_{\ell'}n^1_{\ell''},
\end{equation}
where $\Omega^0_{\ell}=\Omega_\ell\left[\left\{n^0\right\}\right]$, $\Lambda^0_{\ell\ell'}=\eval{\frac{\partial\Omega_\ell[\{n\}]}{\partial n_{\ell'}}}_{n=n^0}$, and $\Gamma^0_{\ell\ell'\ell''}=\eval{\frac{\partial^2\Omega_\ell[\{n\}]}{\partial n_{\ell'}\partial n_{\ell''}}}_{n=n^0}$ are the scattering function and its derivatives with respect to the link occupations evaluated at local equilibrium.

In contrast to these slow, long wavelength dynamics, the magnetic rotation acts directly at the discrete scales of the automaton, since rotation by a large finite angle occurs at each time step. 
Including this magnetic effect in our multi-scale expansion requires keeping the discrete time scale $t_0$ in addition to the slower $t_1$ and $t_2$ scales, so that time and space derivatives are expanded as $\partial_t=\partial_{t_0}+\varepsilon\,\partial_{t_1}+\varepsilon^2\partial_{t_2}$ and $\partial_i=\varepsilon\,\partial_{1i}$, where $\partial_{1i} = \partial/\partial x_{1,i}$.
This procedure, however, will not yield any effect from the magnetic field in the hydrodynamic equations other than the overall rotation of the total momentum.
The fast, short range dynamics of cyclotron orbits entirely decouple from the slow, long-wavelength diffusion of hydrodynamic modes, and in particular viscosities are unchanged from their $B=0$ values.

This can be circumvented by weakening the magnetic effect so that it affects the dynamics at the slower time scale $t_1$. 
This can be done by, for instance, applying the magnetic rotation only once every $1/\varepsilon$ time steps, applying it every step but with probability $\varepsilon$, or rotating the angular variable by $\varepsilon B$ each time step while using $\ell=\lfloor3\theta/\pi\rfloor+1$ to determine the direction of streaming. 
The first two cases involve modifying the dynamics of the automaton and introducing $\varepsilon$ as an explicit parameter of the theory, however due to our replacement of $\ell$ with $\theta$, the latter option can be implemented with a simple rescaling $B=\varepsilon B_1$, where now $B_1=b\pi/3$ is the original (strong) field strength~\footnote{These ``strong'' magnetic fields produces cyclotron motion at the scale of the lattice, which is the same size as the particle mean free path. Since the lattice scale is the smallest length scale present in lattice automata, it is not possible to analyze magnetic fields strong enough to produce cyclotron orbits smaller than the mean free path.}. 

We therefore implement with the Chapman-Enskog multiple scale analysis by substituting
\begin{equation} \label{eq:CE}
\begin{gathered}
    n(\theta,\mathbf{x},t) = n^0(\theta,\mathbf{x},t) +\varepsilon\, n^1(\theta,\mathbf{x},t)\\
    \partial_t = \varepsilon\,\partial_{t_1}+\varepsilon^2\partial_{t_2}, \quad    \partial_i = \varepsilon\,\partial_{1i}, \quad    B= \varepsilon\, B_1
\end{gathered}
\end{equation}
into the Boltzmann equation \ref{eq:Boltzmann2} and expanding up to $O(\varepsilon^2)$.
Collecting terms order by order we obtain
\begin{align}
    O&(1):\,\Omega^0_{\ell}=0 \label{eq:order0}\\
    O&(\varepsilon):\,\mathcal{D}^1_{\ell}\, n^0_{\ell} - \frac{B_1}{3}\epsilon^{ij}\hat{c}_\ell^j\rho u_i = \sum_{\ell'}\Lambda^0_{\ell\ell'}n^1_{\ell'} \label{eq:order1}\\
    O&(\varepsilon^2):\,\partial_{t_2}n^0_\ell+\frac{1}{2}\left(\mathcal{D}^1_{\ell}\right)^2n^0_{\ell}+\mathcal{D}^1_{\ell} n^1_{\ell} -\frac{B_1}{3}\epsilon^{ij}\hat{c}_\ell^j\mathcal{D}^1_{\ell}\rho u_i\nonumber \\ 
    &\qquad\qquad-\frac{B_1^2}{6}\hat{c}_\ell^i\rho u_i=\frac{1}{2}\sum_{\ell',\ell''}\Gamma^0_{\ell\ell'\ell''}n^1_{\ell'}n^1_{\ell''}, \label{eq:order2}
\end{align}
where $\mathcal{D}^1_{\ell}=\partial_{t_1}+\hat{c}_\ell^i\partial_{1i}$ is the material derivative at first order in $\varepsilon$. 
The $O(1)$ equation is a restatement of the definition of equilibrium. 
Summing the $O(\varepsilon)$ and $O(\varepsilon^2)$ equations over $\ell$ directly or after multiplying with a factor of $\hat{\mathbf{c}}_\ell$ gives equations for the dynamics of the density and momentum at these respective orders, given in Appendix \ref{app:CE}. 
Adding these equations with appropriate factors of $\varepsilon$ to reacquire the full spatial and temporal derivatives as in \cref{eq:CE} gives the hydrodynamic equations
\begin{gather}
    \partial_t\rho+\partial_i(\rho u^B_i)=0 \label{eq:continuity}\\
    \partial_t\left(\rho u^B_i\right)= -B\,\epsilon^{ij}\left(\rho u^{B}_j\right) + \partial_j\sigma_{ij}, \label{eq:NavierStokes}
\end{gather}
where 
\begin{multline} \label{eq:stresstensor}
    \sigma_{ij} = -\frac{\rho}{2}\delta^{ij}-\rho\,G(\rho)\sum_\ell\hat{c}_\ell^i\hat{c}_\ell^jQ^{km}_\ell u_ku_m \\
    -\sum_\ell\hat{c}_\ell^i\hat{c}_\ell^j\left[\frac{1}{6}Q_\ell^{km}\partial_k(\rho u_m)+\varepsilon n^1_\ell\right]
\end{multline}
is the stress tensor.
The first of these is the continuity equation, and the second is the Navier-Stokes equation with a Lorentz force term.
The quantity $\rho u^B_i\equiv \left(\delta^{ij}+\frac{B}{2}\epsilon^{ij}\right)\rho u_j$ is the physical momentum that correctly incorporates the perpendicular impulse from the Lorentz force, putting \cref{eq:continuity} into the proper form to reflect local density conservation.
The stress tensor is naturally obtained in terms of $\rho\mathbf{u}$ and not $\rho\mathbf{u}^B$, so after evaluating the contribution from $n^1_\ell$ we must replace $\rho\mathbf{u}$ with the inverse relation $\rho u_i=\frac{1}{1+B^2/4}(\delta_{ij}+\frac{B}{2}\epsilon_{ij})\rho u^B_j$ to identify the proper hydrodynamic stress.
The $O(u^2)$ term in the stress tensor gives advection of the momentum. 

\section{Fluctuations} \label{sec:fluctuations}

The collision-induced fluctuations from equilibrium $n^1_\ell$ can be obtained from \cref{eq:order1}.
First we substitute the explicit form of $n^0_\ell$ up to $O(u)$ to give
\begin{equation}
    \sum_{\ell'}\Lambda^0_{\ell\ell'}n^1_{\ell'} = \frac{1}{3}Q^{ij}_\ell \partial_{1i}(\rho u_j).
\end{equation} 
If the linearized collision matrix $\Lambda^0$ could be inverted then we would immediately obtain $n^1_\ell$ from this equation, but in general $\Lambda^0$ is non-invertible. 
Its eigenvectors and eigenvalues, $\mathbf{v}^\alpha$ and $\lambda_\alpha$ with $\alpha=0,\dots,5$, are the allowed modes of the system and their relaxation rates, so conserved quantities such as density and momentum correspond to modes with $\lambda_\alpha=0$, making the matrix singular. 
To obtain $n^1_\ell$ we must enforce these conservation laws, which can be done by projecting out the null space of $\Lambda^0$ leaving just the modes that are relaxed by the dynamics; these are precisely the modes that should be related to fluctuations via the fluctuation-dissipation theorem. 
The spectral representation naturally gives this projection and allows its inversion,
\begin{gather}
    \Lambda^0 = \widetilde{\Lambda}^0 = \sum_{\alpha}\lambda_\alpha \mathbf{v}^\alpha\otimes\left(\mathbf{v}^\alpha\right)^\dagger \\
    (\widetilde{\Lambda}^0)^{-1}=\sum_{\alpha|\lambda_\alpha\neq0} \frac{1}{\lambda_\alpha}\mathbf{v}^\alpha\otimes\left(\mathbf{v}^\alpha\right)^\dagger,
\end{gather}
where $\widetilde{\Lambda}^0$ represents the projected matrix, and we obtain the general result
\begin{equation} \label{eq:n1general}
    \varepsilon n^1_\ell = \sum_{\ell'}\sum_{\alpha|\lambda_\alpha\neq0}\frac{1}{3\lambda_\alpha}v_\ell^\alpha\left(v^\alpha\right)^\dagger_{\ell'}Q^{ij}_{\ell'}\,\partial_i(\rho u_j) \equiv \widetilde{Q}^{ij}_\ell\,\partial_i(\rho u_j).
\end{equation}
For the hydrodynamics of our chiral automaton we have
\begin{equation} \label{eq:Qtilde}
    \widetilde{Q}^{ij}_\ell=\frac{1}{18}\sum_{r}\left(\frac{e^{2\pi ir/3}}{\lambda}+\frac{e^{-2\pi ir/3}}{\lambda^*}\right)Q^{ij}_{\ell-r},
\end{equation}
where $\lambda = -\tfrac{\rho}{2}(1-\tfrac{\rho}{6})^3[1-i\tfrac{2\sqrt{3}}{3}(p-1/2)]$ and $\lambda^*$ are the eigenvalues of the two modes that contribute to fluctuations, and where $\rho/6=\eval{n^\mathrm{eq}_\ell}_{\mathrm{u}=0}$ is the leading approximation to the equilibrium occupation in the limit of small $u$.
The explicit the form of $\Lambda^0$ and its eigenvalues and eigenvectors yielding this $\widetilde{Q}^{ij}_\ell$ are given in Appendix \ref{app:Fluctuations}. 
The imaginary part of the eigenvalue $\lambda$ is controlled entirely by the chiral term proportional to $p-1/2$. 
As we show below, this imaginary contribution is precisely what generates the Hall-viscosity term in the emergent hydrodynamic equations.

\section{Viscosity} \label{sec:viscosity}

To quantify the effect of the added chiral dynamics we compute the kinematic viscosity tensor $\eta_{ijkl}$, which relates the stress tensor $\sigma_{ij}$ to the gradients of the local fluid momentum field $\rho\mathbf{u}$ as
\begin{equation} 
   \sigma_{ij} = -P\,\delta^{ij}+\eta_{ijkm}\partial_k(\rho u_m),
\end{equation}
where $P$ is the pressure of the fluid. 
The components of the viscosity tensor depend on symmetries of the system, so its form is directly linked of the underlying microscopic dynamics.
Substituting the form of $n^1_\ell$ into the stress tensor obtained for the hydrodynamics of the chiral automaton \cref{eq:stresstensor} we see that the part of the stress tensor proportional to the momentum gradient is
\begin{multline} 
    \sigma'_{ij} = -\sum_{\ell}\hat{c}_\ell^i\hat{c}_\ell^j\left(\frac{1}{6}Q^{km}_{\ell} + \widetilde{Q}^{km}_{\ell}\right)\partial_k(\rho u_m) \\
    =-\sum_{\ell}\hat{c}_\ell^i\hat{c}_\ell^j\left(\frac{1}{6}Q^{kn}_{\ell}+\widetilde{Q}^{kn}_{\ell}\right)\frac{\delta^{nm}+\tfrac{B}{2}\epsilon^{nm}}{1+B^2/4}\partial_k(\rho u^B_m),
\end{multline} 
with the magnetic field entering in the replacement of $\rho\mathbf{u}$ with $\rho\mathbf{u}^B$ as previously noted.
Comparing to the general form of the stress tensor we identify the kinematic viscosity tensor in the presence of the magnetic field
\begin{equation}
    \eta^B_{ijkm} =-\sum_{\ell}\hat{c}_\ell^i\hat{c}_\ell^j\left(\frac{1}{6}Q^{kn}_{\ell}+\widetilde{Q}^{kn}_{\ell}\right)\frac{\delta^{nm}+\tfrac{B}{2}\epsilon^{nm}}{1+B^2/4},
\end{equation}
and see that the effect of the weak magnetic field is to generate new viscosity terms from those acquired for zero field.

With $\widetilde{Q}^{ij}_\ell$ obtained for the FHP automaton with chiral scattering in \cref{eq:Qtilde} we identify the zero-field kinematic shear and Hall viscosities
\begin{gather}
    \eta = \frac{1}{2\rho(1-\rho/6)^3}\frac{1}{1+\frac{4}{3}\left(p-\tfrac{1}{2}\right)^2}-\frac{1}{8} \label{eq:eta} \\
    \eta_H = -\frac{\sqrt{3}}{3\rho(1-\rho/6)^3}\frac{p-\tfrac{1}{2}}{1+\frac{4}{3}\left(p-\tfrac{1}{2}\right)^2} \label{eq:etaH}
\end{gather}
as the coefficients of $\delta^{ik}\delta^{jm}+\delta^{im}\delta^{jk}-\delta^{ij}\delta^{km}$ and $\delta^{ik}\epsilon^{jm}+\epsilon^{ik}\delta^{jm}$ in the full viscosity tensor~\cite{Monteiro2023}.
For nonzero weak magnetic field the corresponding viscosities are written in terms of these,
\begin{equation} \label{eq:viscosityB}
    \begin{pmatrix}\eta^B \\ \eta_H^B \end{pmatrix} = \frac{1}{1+\frac{B^2}{4}}\begin{pmatrix}1 & \frac{B}{2} \\ -\frac{B}{2} & 1 \end{pmatrix}\begin{pmatrix}\eta \\ \eta_H\end{pmatrix}.
\end{equation}
We recover the well-known result for the shear viscosity $\eta$ of the FHP model for $B=0$ and $p=\tfrac{1}{2}$~\cite{FHPReview1987,Henon1987}.
Furthermore, as anticipated chiral scattering and a magnetic field to break parity symmetry are each sufficient to generate a Hall viscosity. 

With maximal parity breaking from scattering alone, i.e. $p=0$ or $1$ with $B=0$, the shear viscosity is minimized and the absolute value of the zero-field Hall viscosity $\eta_H$ is maximized, with their ratio bounded as $1/\sqrt{3}=0.577\leq\abs{\eta_H}/\eta\leq0.732$ as we vary $\rho$, achieving the maximum value for $\rho=3/2$ and the minimum for $\rho\to0,6$, though $\rho\to0$ breaks the assumption that $n^0_\ell\gg \varepsilon \,n^1_\ell$.
For any choice of $p$ and $\rho$, shear viscosity dominates over Hall viscosity for $B=0$, and since we have rescaled the magnetic field strength so that $B\ll 1$ this remains true for all $B$ for which these results apply.

\begin{figure*}[t]
   \centering
   \includegraphics[width=0.9\textwidth]{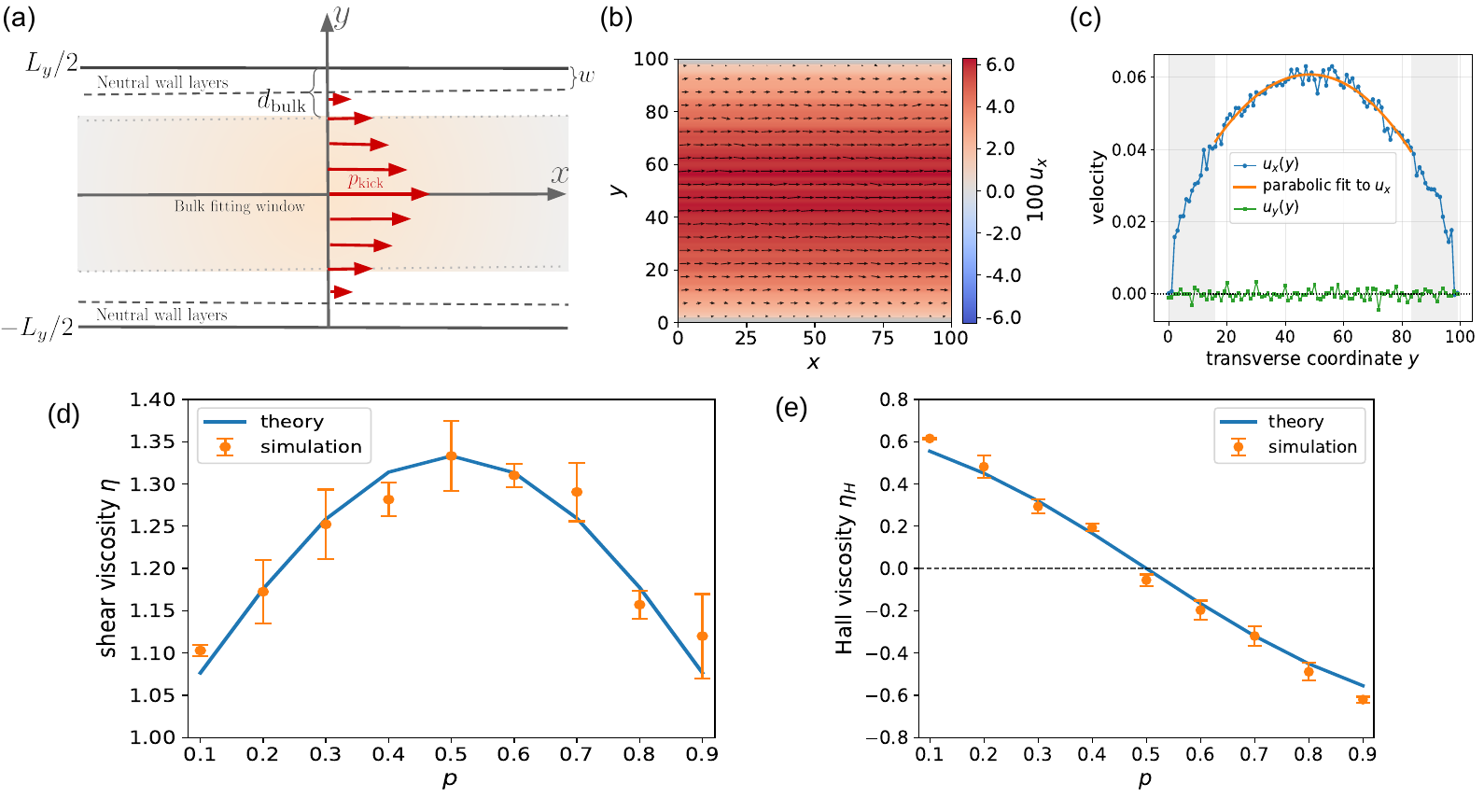}
   \caption{Automaton simulation of chiral hydrodynamic transport.
    (a) Schematic of the Poiseuille geometry used in the simulations. The system is periodic along the longitudinal $x$ direction, with neutral wall layers of thickness $w$ imposed near the upper and lower boundaries. 
    A weak stochastic drive $p_{\rm kick}$ injects longitudinal momentum in the bulk, while viscosity extraction is performed in the central fitting window after excluding boundary-adjacent regions of thickness $d_{\rm bulk}$.
    (b) Time-averaged flow at $p=0.7$, with color indicating $100u_x$ and arrows showing the coarse-grained velocity field.
    (c) Velocity profiles $u_x(y)$ and $u_y(y)$; the solid curve shows the quadratic bulk fit to $u_x(y)$.
    (d,e) Shear viscosity $\eta$ and Hall viscosity $\eta_H$ versus chirality parameter $p$.
    Blue curves denote analytical predictions, and orange symbols denote calibrated simulation results with error bars.
    All simulations are run for $B=0$. 
    }
    \label{fig:ZeroFieldSim}
\end{figure*}

\section{Automaton Simulations} \label{sec:simulation}

We test the hydrodynamic predictions through direct numerical simulations of the chiral FHP automaton in a pipe-flow geometry, also known as Poiseuille flow~\cite{Kadanoff1987,Kadanoff1989}. 
The simulation evolves Boolean occupations $N_\ell(\mathbf{x},t)\in\{0,1\}$ on a finite triangular lattice, with each time step consisting of the collision and streaming updates defined in \cref{sec:automaton}. 
From the particle-level dynamics, we compute coarse-grained density, momentum, velocity, and stress fields for comparison with the analytical viscosities of \cref{sec:viscosity}.

The simulated Poiseuille system consists of a lattice of dimensions $L_x=L_y=100$ with periodic boundary conditions taken in the $x$ direction. 
A weak stochastic drive is applied to the particles, flipping left-moving channels to right-moving channels with probability $p_\mathrm{kick}=2.5\times10^{-4}$ when allowed by occupation rules, except within neutral wall layers of thickness $w=2$ near the top and bottom boundaries to implement no-slip boundary conditions. 
This injects momentum along $x$ while preserving particle number, producing an overall flow.
A schematic of this geometry is shown in \cref{fig:ZeroFieldSim}(a).
 
For each simulation of this Poiseuille system we fix a value of the chiral scattering parameter $p$, varied between $0.1$ and $0.9$ across different simulations. 
The system is initialized at half filling by randomly choosing, at each site, one of the two zero-momentum three-particle configurations, $C_{3,1}$ or $C_{3,2}$, giving half-filling $\rho=3$ with no imposed initial flow. 
Simulations are then run for $3\times10^4$ time steps, with the first $10^4$ discarded as warm-up, and density and velocity profiles sampled every 10 steps thereafter. 
The resulting hydrodynamic fields are averaged over time and along $x$, giving one-dimensional profiles $\rho(y)$, $u_x(y)$, $u_y(y)$, and stress components.  
When extracting properties of the fluid from these density and velocity profiles, fits are restricted to a central bulk window in order to reduce boundary effects, excluding boundary regions of thickness $d_{\rm bulk}=14,16,18$.

The time-averaged velocity field shown in \cref{fig:ZeroFieldSim}(b) confirms that this driven chiral FHP automaton develops a stable Poiseuille-like response dominated by flow along $x$, and \cref{fig:ZeroFieldSim}(c) shows the one-dimensional velocity profiles: $u_x(y)$ follows the expected quadratic bulk form~\cite{Kadanoff1987}, while $u_y(y)$ remains small. 
The shear viscosity is extracted from the curvature of the Poiseuille velocity profile, following the standard lattice gas hydrodynamic interpretation of viscous momentum transport~\cite{FHPReview1987,Henon1987,Kadanoff1987}. 
In the bulk region, the averaged longitudinal velocity is fit to $u_x(y)=a_2y^2+a_1y+a_0$, and the weak-drive hydrodynamic balance gives $\partial_y^2u_x=2a_2$. 
We therefore extract the Poiseuille shear viscosity estimate as 
$\eta_{\rm P}=f_x/|2a_2|$, where $f_x$ denotes the coarse-grained body-force density associated with the imposed stochastic drive.
Microscopically, this drive is controlled by the kick probability $p_{\rm kick}$: each accepted kick converts a left-moving occupation into a right-moving one and injects a fixed amount of longitudinal momentum.
Because the accepted kick rate depends on local exclusion constraints, $f_x$ is best viewed as the effective momentum-injection scale of the drive, which fixes the overall normalization of the extracted Poiseuille viscosity.

The Hall viscosity is extracted from the parity-odd stress response in the same steady state, following the standard odd-viscous stress structure in two-dimensional parity-breaking fluids~\cite{Avron1995,Banerjee2017,Monteiro2023}. 
For this purpose, the simulation stress tensor is defined as the non-convective part of the kinetic momentum flux,
$\sigma_{ij}(y)=\Pi_{ij}(y)-\rho(y)u_i(y)u_j(y)$, 
with the same sign convention used in the analytical viscosity formulas. 
Specifically, we compute $\Delta\Sigma(y)=\sigma_{xx}(y)-\sigma_{yy}(y)$ and fit it against $\partial_y[\rho(y)u_x(y)]$ over the same bulk window, with the slope determining $\eta_{H,{\rm P}}=-(1/2)\Delta\Sigma/\partial_y(\rho u_x)$.

In \cref{fig:ZeroFieldSim}(d) and (e) we compare the normalized simulation estimates with the analytical predictions for $\eta$ and $\eta_H$ as functions of the chirality parameter $p$. 
The simulations capture both the nonmonotonic dependence of the shear viscosity and the sign-changing Hall response, showing that the microscopic chiral collision rules reproduce the expected zero-field odd-viscous hydrodynamics.
For each $p$, results are averaged over ten random seeds, and the error bars combine seed-to-seed fluctuations with the systematic spread from varying the bulk fitting window size $d_\mathrm{bulk}$, added in quadrature. 
The Poiseuille estimates are reported with one overall normalization, and the same normalization is then used for the entire chirality sweep. 
Thus, the simulation comparison focuses on the predicted dependence on $p$, including the nonmonotonic shear response and the sign-changing Hall viscosity.

\section{Discussion and Future directions} \label{sec:discussion}
In this work, we have developed a chiral generalization of the FHP lattice-gas automaton.
Chirality is encoded both through asymmetric two-body collision rules and through a systematic rotation of particle velocities that mimics the effect of an external magnetic field. 
After coarse-graining the automaton dynamics and performing a Chapman-Enskog expansion we derived the corresponding continuum hydrodynamics.

We find that chirality suppresses the shear viscosity and generates an additional transport coefficient: the Hall viscosity. 
This emergent Hall viscosity is consistent with the breaking of parity while preserving the isotropy of the continuum theory. 
We obtain exact analytical expressions for the shear and Hall viscosity as functions of the density and the chiral parameters. 
To test these predictions, we simulate Poiseuille flow in the chiral FHP automaton and find excellent agreement with the analytically derived transport coefficients. 
Our results demonstrate that lattice-gas automata provide a controlled framework for bridging microscopic chiral dynamics and macroscopic odd-fluid behavior. 

This work also opens several natural directions for future study.  
One application of this approach is to electron hydrodynamics, where it provides a way to investigate whether encoded or explicitly broken microscopic symmetries can serve as effective proxies for Fermi-surface geometry or for the presence of an external magnetic field. 
The automaton rules can also be generalized to incorporate activity within the lattice framework. 
For example, it would be interesting to construct lattice versions of Vicsek-type automata that exhibit flocking~\cite{Vicsek1995, TonerTu1995}, and to study the interplay between chirality, symmetry breaking, and collective flocking dynamics.

Another promising direction is to exploit the Boolean structure of cellular automata to quantize the dynamics as quantum circuits, as discussed in Ref.~\cite{Singh2025}. 
In the present setting, parity-breaking automaton rules may induce chiral propagation of quantum information, suggesting a possible route to engineering directional information flow in quantum computing architectures. This direction is currently under investigation and will be addressed in a separate publication.

\section*{Acknowledgments}
This work was supported by National Science Foundation (NSF) Grant No.~DMR-2315063 (A.A., P.G., and S.G.), NSF Grant No.~DMR-2315064 (A.R.), and NSF Grant No.~DMR-2611305 (T.I.). S.G. was supported in part by grant NSF PHY-2309135 to the Kavli Institute for Theoretical Physics (KITP), where part of this work was carried out. S.G. would also like to thank Kranthi K. Mandadapu for useful discussions.

\appendix

\section{Equilibrium Solution} \label{app:equilibrium}
The equilibrium link occupation $n^\mathrm{eq}_\ell$ is determined as the function that maximizes the entropy at each site.
This entropy is
\begin{multline}
    S = \sum_\ell\left[n_\ell\log n_\ell + (1-n_\ell)\log(1-n_\ell)\right] \\
    + h(\rho,\mathbf{u})\sum_\ell n_\ell + \mathbf{q}(\rho,\mathbf{u})\cdot\sum_\ell\hat{\mathbf{c}}_\ell n_\ell,
\end{multline}
where the Lagrange multipliers $h(\rho,\mathbf{u})$ and $\mathbf{q}(\rho,\mathbf{u})$ depend on the density and velocity and enforce local conservation of these quantities. 
Setting $\partial S/\partial n_\ell=0$ and solving for the occupation function gives
\begin{equation}
    n^\mathrm{eq}_\ell = \frac{1}{1+e^{-(h(\rho,\mathbf{u})+\mathbf{q}(\rho,\mathbf{u})\cdot\hat{\mathbf{c}}_\ell)}},
\end{equation}
the Fermi-Dirac distribution. 
For a system with more locally conserved quantities, each would appear as a new term in the exponential.

The form of the small-$u$ expansion of this equilibrium solution is obtained by first expanding the undetermined functions $h$ and $\mathbf{q}$ up to second order in $\mathbf{u}$ subject to the constraints imposed by inversion symmetry of the system--$h(\rho,\mathbf{u}) = h(\rho,-\mathbf{u})$ and $\mathbf{q}(\rho,\mathbf{u}) = -\mathbf{q}(\rho,-\mathbf{u})$.
This requires that the expansion of $h$ contain only even powers of $\mathbf{u}$, $h\approx h_0(\rho) + \tfrac{1}{2}h^{ij}_2(\rho)u_iu_j$ and $\mathbf{q}$ only odd powers, $\mathbf{q}\approx \mathbf{q}_{1}^i(\rho)u_i$, leaving three undetermined functions $h_0, h_2^{ij},$ and $\mathbf{q}_1^i$.
Expanding the full $n^\mathrm{eq}_\ell$ up to second order in $\mathbf{u}$ and requiring the density $\rho$ and momentum $\rho\mathbf{u}$ be obtained by \cref{eq:uniform} determines the form of these functions giving \cref{eq:equilibrium}. 

\section{Chapman-Enskog Expansion} \label{app:CE}
With substitution of the expansions \cref{eq:CE} into the Boltzmann equation \cref{eq:Boltzmann2} we expand in powers of $\varepsilon$, keeping up to second order, giving the three equations \cref{eq:order0,eq:order1,eq:order2}.
The first is a restatement of the equilibrium condition and plays no further role in the analysis.
Summing the remaining two over $\ell$ directly, or first multiplying by $\hat{c}_\ell^i$ then summing over $\ell$ we obtain
\begin{align}
    O(\varepsilon)& \begin{cases}
        \partial_{t_1}\rho +\partial_{1i}\left(\rho u_i \right) =0 \\
        \partial_{t_1}(\rho u_i) = -B_1\,\epsilon^{ij}\rho u_j +\partial_{1j}\Pi^0_{ij}
    \end{cases}\\
    \nonumber \\
    O(\varepsilon^2)& \begin{cases}
        \partial_{t_2}\rho + \frac{B_1}{2}\epsilon^{ij}\partial_{1i}(\rho u_j) =0\\
        \partial_{t_2}(\rho u_i) = -\frac{B_1}{2}\epsilon^{ij}\partial_{t_1}(\rho u_j)-\frac{B_1^2}{2}\rho u_i + \partial_{1j}\Pi^1_{ij},
    \end{cases}
\end{align}
where $\Pi^0$ and $\Pi^1$ are the leading and first-order terms of the momentum-flux tensor
\begin{align}
    \Pi^0_{ij} &= -\frac{\rho}{2}\delta^{ij}-\rho\,G(\rho)\,\sum_\ell \hat{c}_\ell^i\hat{c}_\ell^jQ^{km}_{\ell}u_k u_m \label{eq:Pi0}\\
    \Pi^1_{ij} &=-\sum_\ell\hat{c}_\ell^i\hat{c}_\ell^j \left[n^1_{\ell} + \frac{1}{6}Q^{km}_\ell\,\partial_{1k}(\rho u_m)\right], \label{eq:Pi1}
\end{align} 
which sum to give the full stress tensor. 
We have also used here that the vanishing of the collision factor under these sums, \cref{eq:conservation}, reflecting particle number and momentum conservation, implies that the sum over the expansion of this quantity vanishes at each order in $\varepsilon$, i.e. $\sum_{\ell,\ell'}\Lambda^0_{\ell\ell'}n^1_{\ell'}=0$, and so on. 

The $O(\epsilon^2)$ equations are obtained by simplifications arising from the equations obtained at $O(\epsilon)$. 
Multiplying these equations by appropriate factors of $\varepsilon$ and adding them together gives the hydrodynamic equations \cref{eq:continuity,eq:NavierStokes} after using \cref{eq:CE} to recover the derivatives with respect to $t$ and $\mathbf{x}$ and the magnetic field strength $B$.

\section{Fluctuations} \label{app:Fluctuations}
For the hydrodynamics obtained from our chiral automaton the linearized collision matrix evaluated in equilibrium at lowest order in the velocity $u$ is
\begin{widetext}
\begin{equation}
    \Lambda^0 = \begin{pmatrix} 
        -\gamma-\beta & \gamma(1-p)+\beta & p\gamma-\beta & -\gamma+\beta & \gamma(1-p)-\beta & p\gamma+\beta \\
        p\gamma+\beta & -\gamma-\beta & \gamma(1-p)+\beta & p\gamma-\beta & -\gamma+\beta & \gamma(1-p)-\beta \\
        \gamma(1-p)-\beta & p\gamma+\beta & -\gamma-\beta & \gamma(1-p)+\beta & p\gamma-\beta & -\gamma+\beta \\
        -\gamma+\beta & \gamma(1-p)-\beta & p\gamma+\beta & -\gamma-\beta & \gamma(1-p)+\beta & p\gamma-\beta \\
        p\gamma-\beta & -\gamma+\beta & \gamma(1-p)-\beta & p\gamma+\beta & -\gamma-\beta & \gamma(1-p)+\beta \\
        \gamma(1-p)+\beta & p\gamma-\beta & -\gamma+\beta & \gamma(1-p)-\beta & p\gamma+\beta & -\gamma-\beta 
    \end{pmatrix},
\end{equation}
\end{widetext}
in terms of $\gamma = (\rho/6)(1-\rho/6)^3$ and $\beta = (\rho/6)^2(1-\rho/6)^2$, since $\rho/6$ is the leading approximation to $n^\mathrm{eq}_\ell$ in the limit of low Mach number. 
This is a circulant matrix, so the components of its eigenvectors can be immediately written as $v^\alpha_\ell = \omega^{\alpha(\ell-1)}/\sqrt{6}$ with $\omega = e^{i\pi/3}$ for $\alpha=0,\dots,5$ and $\ell=1,\dots,6$. 
As expected from conservation of density and two components of momentum, three of these vectors have zero eigenvalue, $\lambda_0=\lambda_1=\lambda_5=0$.
The remaining modes have eigenvalues $\lambda_3= -6\beta$ and $\lambda_2=(\lambda_4)^* = -3\gamma+i2\sqrt{3}(p-\tfrac{1}{2})\,\gamma$; breaking parity symmetry $p\neq1/2$ causes two to become complex. 
Because of the lattice's inversion symmetry we find $\sum_{\ell}(v^3)^\dagger_\ell Q^{ij}_\ell = 0$, so only the two modes $\alpha=2,4$ contribute to the fluctuations giving
\begin{multline}
    n^1_\ell = \frac{1}{18}\sum_{\ell'}\left(\frac{e^{2\pi i(\ell-\ell')/3}}{\lambda_2}+\frac{e^{-2\pi i(\ell-\ell')/3}}{\lambda_2^*}\right)Q^{ij}_{\ell'}\,\partial_i(\rho u_j).
\end{multline}\\

\bibliography{references.bib}
\end{document}